\documentclass{PoS}

\newcommand{\beq}{\begin{eqnarray}}
\newcommand{\eeq}{\end{eqnarray}}

\newcommand{\real}{{\sf I}\kern-.12em{\sf R}}
\newcommand{\comp}{{\sf I}\kern-.50em{\sf C}}
\newcommand{\unity}{{\sf I}\kern-.54em{\sf 1}}
\def\spose#1{\hbox to 0pt{#1\hss}}
\def\ltapprox{\mathrel{\spose{\lower 3pt\hbox{$\mathchar"218$}}
 \raise 2.0pt\hbox{$\mathchar"13C$}}}

\title{Lattice QCD with purely imaginary sources at zero and non-zero temperature.}

\ShortTitle{Lattice QCD with purely imaginary sources at zero and non-zero temperature.}

\author{Massimo D'Elia\\
\\
Dipartimento di Fisica dell'Universit\`a
di Pisa and INFN - Sezione di Pisa,\\ Largo Pontecorvo 3, I-56127 Pisa, Italy\\
        E-mail: \email{delia@df.unipi.it}}

\abstract{
We discuss various aspects and recent progress 
concerning lattice QCD studies in the presence of external sources.
We focus, in particular, on issues regarding QCD 
with non-zero imaginary chemical potentials or with a
$\theta$-term, and on the properties of strongly interacting matter
in the presence of electromagnetic background fields.}

\FullConference{The 32nd International Symposium on Lattice Field Theory\\
23-28 June, 2014\\
Columbia University New York, NY}

\begin{document}

\section{Introduction}

Exploring a quantum field theory in the presence of external sources
is a common
way to investigate its properties.
A relevant example
is the response to chemical potentials and background fields. 
We shall consider a class of sources which, in the Euclidean formulation,
maintain the positivity of the path-integral formulation of the 
partition function, which for a non-Abelian lattice gauge theory
is written as
\beq
Z = 
 \int {\cal D}U {\cal D}\psi {\cal D}\bar{\psi}
e^{-( S_G[U] + \bar\psi M[U] \psi)} = 
\int  {\cal D}U e^{-S_G[U]} \det M[U]\, ,
\eeq
so that their effect is directly explorable
by numerical Monte-Carlo simulations, 
without the need to deal with a difficult sign problem.

In fact, such kind of sources are often considered in order to partially circumvent the sign problem which is met in the presence of real sources. 
The best known example
is QCD in the presence of a baryon 
chemical potential $\mu_B$, whose partition function 
is written as 
\begin{equation}
Z(T,\mu_B) = {\rm Tr} \left( e^{-\frac{H_{\rm QCD} - \mu_B N_B}{T}}
\right) \, . 
\end{equation}
In the path-integral formulation a non-zero $\mu_B$ amounts 
to a modification of the quark action density\ 
%\begin{equation}
$
\bar{\psi} \left( m + \gamma_\nu D_\nu \right) \psi \to   
 \bar{\psi} \left( m + \gamma_\nu (\partial_\nu + i (A_\nu - i\, \delta_{0\nu}\, \mu) \right) \psi\  
$
%\end{equation}
where $\mu = \mu_B/3$ is the quark chemical potential. If $\mu$ is real, 
the 
anti-hermiticity properties of $\gamma_\nu D_\nu$ are lost
and the fermion determinant $\det M(\mu)$ becomes complex (sign problem),
so that Monte-Carlo simulations cannot be used, unless 
opposite chemical potentials are given to 
up and down quarks (isospin chemical potential) or to 
left-handed and right-handed quarks (axial chemical
potential $\mu_5$).

Approximate solutions consist in computing $\mu_B$ derivatives 
at $\mu_B = 0$ (i.e. the cumulants of the baryon number distribution,
or generalized susceptibilities), in order to reconstruct
the free energy dependence as a Taylor expansion in $\mu_B$~\cite{tay},
or in performing importance sampling by taking the positive part of the path-integral measure
and then reweighting gauge configurations with the complex factor~\cite{rew}.
Both methods have significant limitations: reweighting in general fails 
as the statistics needed to correctly sample the partition function
grow exponentially with the volume; 
Taylor expansion is limited to 
the small $\mu_B$ region, since the computational effort needed to determine
the terms of the series grows very rapidly with the order of expansion.

An alternative is to introduce a purely imaginary quark 
chemical potential, $\mu = i \mu_I$~\cite{various_im,critsurf,ceavari,mdfs09,CGFMRW,OPRW,PP_wilson,alexandru,wumeng,ccp,nf2-order,Nagata:2014fra,corvo,takaha}: in this case
the $\gamma_5$-hermiticity of the fermion matrix is preserved
and the quark determinant is real and positive. In the standard way 
to introduce a chemical potential on the lattice~\cite{khg}, that amounts
to adding a constant 
 $U(1)$ phase to temporal links appearing in the Dirac operator,
$ U_t(n) \to e^{i a \mu_I} U_t(n)$, i.e. it is equivalent
to a twist by an angle $\theta_I = \mu_I/T$ in the fermionic temporal 
boundary  conditions.
Charge conjugation implies that the partition function 
is even in $\mu$, so that switching to imaginary 
$\mu$ is actually like sending
$\mu^2 \to -\mu^2$, and analytic continuation can be performed
in the $T - \mu^2$ plane, when possible,
to infer properties at real $\mu$ from simulations at imaginary $\mu$.

The partition function at imaginary $\mu$ contains further interesting 
information. $Z(\theta_I)$ is of course periodic in $\theta_I$, with period 
$2 \pi$, but the additional symmetry of the path-integral measure
and of the pure gauge term under center $Z_{N_c}$ transformations,
where $N_c$ is the number of colors, implies that 
the partition function is actually $2 \pi/N_c$ periodic in $\theta_I$, i.e.
$2 \pi$ periodic in $\theta_B \equiv {\rm Im} (\mu_B)/T$~\cite{rw}.
Such a periodicity is directly linked to the fact that only globally
colorless 
physical states (hence with an integer baryon number $n_B$) contribute
to the partition function, both in the low $T$ and 
in the high $T$ phase, however it is differently realized, 
smoothly or with phase transitions at $\theta_B = \pi$ or
odd multiples of it (known as Roberge-Weiss transitions~\cite{rw}),
depending on the effective degrees of freedom
of the theory, which are hadrons below $T_c$ and quarks and 
gluons above. Lattice simulations have provided plenty of evidence
of this different behavior from direct simulations
at imaginary chemical potentials; moreover, the extraction of 
the Fourier coefficients of the free energy, 
considered as a periodic function of $\theta_B$, gives
direct access to the canonical partition function, i.e. 
at fixed baryon number~\cite{cano}. 
\\

A closely analogous example is given by non-Abelian gauge
theories, with or without dynamical fermions, with a topological
$\theta$ term, 
\beq
Z(\theta) = 
\int  {\cal D}U e^{-S_G[U] + i \theta Q[U]} \det M[U]
\eeq
where 
$Q = \int d^4 x \frac{g_0^2}{64\pi^2} 
\epsilon_{\mu\nu\rho\sigma} G_{\mu\nu}^a(x) 
G_{\rho\sigma}^a(x)$
is the topological charge, which enters various aspects
of CP-violation in strong interactions.
Also in this case a non-zero $\theta$ makes 
the path-integral measure complex, thus hindering direct 
Monte-Carlo simulations.
One is usually interested in the $\theta$-dependence of various
quantities of physical interest, including the vacuum energy
and, at finite $T$, the free energy density, whose essential
features can be reconstructed, in a Taylor expansion approach, in terms
of the cumulants of the probability 
distribution of $Q$ at $\theta = 0$.
The introduction of a purely imaginary $\theta$ has been 
considered as an alternative since long~\cite{bhanot},
with many interesting results obtained in the last few years
for QCD and QCD-like theories~\cite{thetaim,takuedm,aoki_1,dene,guo}.

There is an interesting analogy between the physics at non-zero, real
$\theta$ and that at non-zero, imaginary $\mu$, which is linked to the 
$2 \pi$ periodicity in $\theta$ deriving from the fact that $Q$ is integer-valued.
The analogy involves a sort of duality: the periodicity is non-trivially
realized in the low $T$ phase of the theory, where the actual dependence is on 
$\theta/N_c$ and phase transitions appear at $\theta = \pi$
or odd multiples
of it, while a smooth periodic behavior is realized in the deconfined phase.
However, in this case the periodic behavior is on the same
side where a sign problem occurs, so that its different
realizations at high $T$ or low $T$ cannot 
be checked by direct simulations, as for imaginary $\mu$,
but only indirectly, through the determination
of the cumulants~\cite{bdpv}.
\\

Electromagnetic sources 
have been considered since the early days of lattice simulations
to probe the electromagnetic properties of hadrons~\cite{hadron}.
An electromagnetic 
background field $a_\mu$ modifies the covariant derivative, for 
a quark with electric charge $q$, as follows:
\beq
D_\mu = \partial_\mu + i\, g A^a_\mu T^a  \ \ \  \to \ \ \  \partial_\mu + i\, g A^a_\mu T^a + i\, q a_\mu \, .
\eeq
The $\gamma_5$-hermiticity of the Dirac operator, hence the 
positivity of the quark determinant, is preserved 
only if $a_\mu$ is real valued: on the lattice,
$SU(3)$ links appearing in the fermion matrix
get multiplied by a pure phase factor, 
$U_\mu(n) \to U_\mu(n) u_\mu(n)$ where
$u_\mu (n) \simeq \exp(i\, q\, a_\mu(n)) \in U(1)$, which is 
similar to adding an imaginary chemical potential. 
In the Euclidean path-integral formulation that corresponds 
to a non-zero imaginary electric field, while real electric
fields
lead to a sign problem. Numerical studies regarding
the electric polarizabilities of hadrons usually involve
such imaginary electric fields: the ground state energy shift, which 
at the lowest order is proportional to $|\mathbf E|^2$, is analitically continued 
from imaginary to real electric fields in order to determine the
polarizability.

Magnetic fields, instead, are real, 
so that numerical simulations can be performed 
without any technical obstruction. That has opened the way
to extensive simulations of QCD in the presence of magnetic
background fields, whose interest
is justified by the many phenomenological contexts where
they may play a role, including
the physics of 
compact astrophysical
objects like magnetars~\cite{magnetars}, 
of non-central heavy ion collisions~\cite{hi} 
and of the early Universe~\cite{vachagrarub}, which involve
magnetic fields
going from $10^{10}$ Tesla up to $10^{16}$ 
Tesla ($|e \mathbf B| \sim 1$ GeV$^2$).
\\

Quantum Field Theories formulated in moving frames,
e.g., Lorentz boosted or rotating, 
can also be viewed as systems with 
external sources, coupled respectively to the
total momentum or to the angular momentum of the system.
For instance, the partition function in a frame moving with speed
$\mathbf v$ is written as
\beq
Z(T,\mathbf v) = {\rm Tr}\left( e^{-(H - \mathbf v \cdot \mathbf P)/T} \right)
\label{movings}
\eeq
where $\mathbf P$ is the total momentum of the system. However, the 
path-integral representation of Eq.~(\ref{movings}) leads 
to a sign problem, since the term coupled to $\mathbf v$ in the
Euclidean action is purely imaginary, as can be easily verified even
in the simplest case of the
Euclidean path-integral for a non-relativistic particle
after a Galileo transformation. Also in this case, the sign 
problem can be avoided by considering the speed
$\mathbf v$ as a purely imaginary quantity, i.e. 
$\mathbf v = i \mathbf \xi$, so that
\beq
Z(T,\mathbf v) = {\rm Tr}\left( e^{-(H - \mathbf v \cdot \mathbf P)/T} \right)
\to {\rm Tr}\left( e^{-(H - i \mathbf \xi \cdot \mathbf P)/T} \right) = 
\sum_n \langle n | e^{-(H - i \mathbf \xi \cdot \mathbf P)/T} | n \rangle \, .
\label{moving2}
\eeq
That, on its turn, can be viewed, for systems which are 
translationally invariant, as a modification of the standard thermodynamical
trace, consisting in applying 
the translation operator $\exp(i \mathbf \xi \cdot \mathbf P)/T$ 
to the right-hand side eigenstate $| n \rangle$. That
has a representation in terms of a path-integral
with spatially shifted,
instead of standard periodic (or anti-periodic),
boundary conditions in the time direction  ($\mathbf \xi$ being the shift).
Numerical simulations of such system are interesting, since the
$\mathbf \xi$-dependence of the theory can be used to 
derive the equation of state~\cite{xi}.

A similar continuation, from real to imaginary angular velocities,
permits the numerical simulations of QCD in a rotating frame~\cite{rot1}, which
in this case can be useful and interesting by itself, because
of the phenomenology related to heavy ion collisions and to
compact stars.
\\

In the following I will cover in more detail only some of the recent 
developments concerning the topics above. 
In cases where a sign problem exist, many efforts 
are being dedicated to the exploration of new approaches, 
which could eventually lead to an effective solution. Such efforts 
include Langevin simulations for generic complex actions~\cite{langevin},
lattice simulations on a Lefschetz thimble~\cite{timble}, 
density of states methods~\cite{densitystates},
formulations in terms of dual variables~\cite{dualg},
tensor Renormalization Group techniques~\cite{trn}, 
effective Polyakov loop models~\cite{effective,effective2}
and many other approaches, which 
unfortunately will not be discussed in this context.

\section{Imaginary chemical potentials}

The location of the (pseudo)critical temperature $T_c$ 
and the nature of the transition as a function of $\mu$ 
are two key issues for the QCD phase diagram. Determining 
$T_c(\mu)$ for small values of $\mu$ is a well defined 
goal for analytic continuation:
$T_c$ is located for different values
of $\mu_I$ and then it is fitted to a  truncated 
Taylor series or to other expansions 
(e.g., Pad\'e~\cite{mpl} approximants) and continued to real $\mu$.
The reliability of the determination of the quadratic term, i.e. the curvature
\beq
\frac{T_c(\mu_B)}{T_c}=1-\kappa \left(\frac{\mu_B}{T_c}\right)^2\, +\, 
O(\mu_B^4)\, ,
\label{curvature}
\eeq
has been verified in many different ways: by direct comparison 
with real $\mu$ simulations for theories without a sign problem~\cite{ceavari}
and by
comparison with the outcome from other approaches, like
reweighting~\cite{fkrev} or Taylor expansion~\cite{laermann}.
Determining non-linear contributions in $\mu^2$
is harder~\cite{ceavari} and faces a typical problem
for analytic continuation: contributions which are suppressed
for $\mu^2 < 0$ may become important for $\mu^2 > 0$, because of
the different signs in the Taylor expansion.

\begin{figure} 
\begin{center}
\includegraphics*[width=0.62\textwidth]{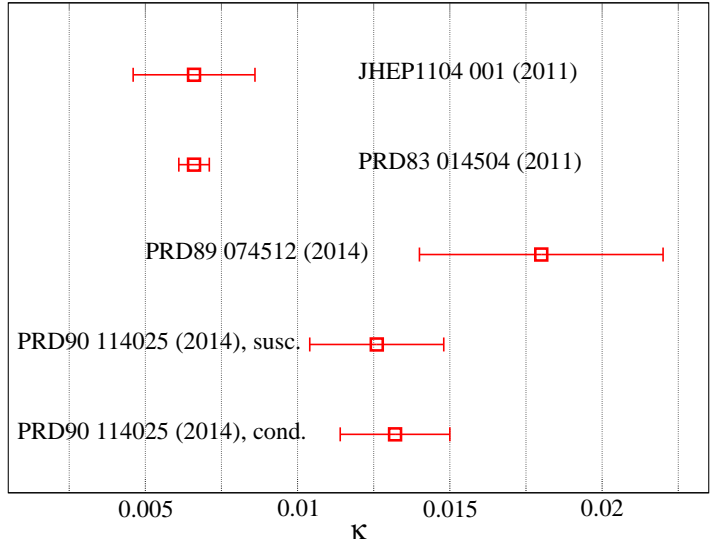}
\caption{
Curvature of the pseudo-critical line from
different studies.
From bottom to top: 
 \emph{i}) and \emph{ii}) analytic continuation, chiral condensate
and chiral susceptibility, stout staggered quarks~\cite{corvo};
 \emph{iii}) analytic continuation, chiral susceptibility with $\mu_s = \mu_l$, 
HISQ staggered quarks~\cite{ccp}; 
 \emph{iv}) Taylor expansion, chiral susceptibility, staggered quark p4-action~\cite{Kaczmarek2011}; 
 \emph{v}) Taylor expansion, chiral condensate, stout staggered quarks~\cite{Endrodi2011}.}
\label{fig1} 
\end{center}
\end{figure}

Recently, such studies have been extended to realistic discretizations
of $N_f = 2+1$ QCD with physical or almost physical quark masses
in two different studies, the first adopting HISQ 
staggered quarks~\cite{ccp}, the other adopting 
stout staggered quarks~\cite{corvo}. In both cases 
a tree level Symanzik improved gauge action and lattices
with $N_t = 6$ and $N_t = 8$ have been adopted. Results for the curvature
$\kappa$, Eq.~(\ref{curvature}),
are reported in Fig.~\ref{fig1} and compared to previous determinations
obtained by Taylor expansion: analytic continuation brings 
consistently larger results, with a discrepancy which is presently 
at the 2-$\sigma$ level when systematic effects are taken into account~\cite{corvo}.
A continuum extrapolation of these results is surely needed to clarify
the issue, in view of a definite comparison with heavy ion phenomenology.

Like other methods to avoid the sign problem, 
analytic continuation will hardly say something definite
regarding the possibile presence of a critical endpoint 
along the (pseudo)critical line, where the transition turns
from crossover to first order.
Numerical results show
that the introduction of an imaginary $\mu$ tends to enlarge the
region of small quark masses in the Columbia plot
 where the chiral transition is first order~\cite{critsurf}; then, by analytic continuation
(of a critical surface, in this case), a real $\mu$
would shrink the same region, thus disfavoring the appearance 
of a first order at the physical point. 
A safe conclusion, at present, is that the critical point,
if any, is not to be found in the small $\mu$ region, and may 
not be directly connected to the chiral first order region taking place
at $\mu = 0$. That diminishes the chances to catch it 
before a complete solution to the sign problem is reached.

The phase structure at imaginary $\mu$ presents many interesting features by itself:
the Roberge-Weiss first order transition lines, which are present 
in the high-$T$ region and for $\theta_B = {\rm Im}(\mu_B)/T = n \pi$,
must stop with critical endpoints at some critical temperature $T_{RW}$,
in order to match with the smooth periodic structure at low $T$.
These are the only critical points 
in the QCD phase diagram 
whose existence can be predicted for sure (even if on the unphysical side); they also coincide, 
by exchange of the temporal direction with a spatial one,
 with the finite size transition at which charge conjugation symmetry breaks~\cite{finitesize,CGFMRW}.
Recent studies, performed both with staggered~\cite{CGFMRW,OPRW} 
and Wilson fermions~\cite{PP_wilson,alexandru,wumeng} for $N_f = 2$ and $N_f = 3$, 
have shown that such endpoints
turn from first to second order, and then to first order again, as the 
quark masses are tuned from zero to infinity, with tricritical points separating the different regions.
The evolution of this phase structure can be monitored from ${\rm Im}(\mu_B)/T = \pi$ down 
to $\mu_B = 0$, and the use of tricritical scaling~\cite{OPRW} can be particularly helpful
to extrapolate results to the chiral limit. That has permitted to 
make progress on a long-standing issue, like the order of the chiral transition
for $N_f = 2$ QCD in the massless limit~\cite{nf2-order}: the first order region
which is present at ${\rm Im}(\mu_B)/T = \pi$ remains
finite down to $\mu_B = 0$, in agreement with previous lattice evidence~\cite{nf2-pisa}.
Given the coarse lattices ($N_t = 4$ with standard staggered fermions), such a result
is preliminary, but shows that the strategy is promising for future improved studies.

The non-trivial phase structure at imaginary $\mu$ has given rise to many model 
studies~\cite{model-rw} and to further speculations
about possible phenomena related to it~\cite{Nagata:2014fra}.
An interesting aspect that we would like to stress is the possibility 
to test lattice methods, currently used to locate the critical endpoint (like looking
at the radius of convergence of the Taylor series), on the critical points
present at imaginary $\mu$, whose location is well known from direct simulations; 
this is something that should be done in the future, in order to better
assess present systematic uncertainties on the determination of the critical point.

Simulations at imaginary $\mu$ provide useful information for many effective models used to explore
the QCD phase diagram: in the same spirit of analytic continuation, 
the effective models are continued from real to imaginary chemical 
potentials, where the free parameters of the model are fixed 
against the results of lattice simulations. Examples 
are given by the effective Polyakov line action at finite
chemical potential~\cite{effective}, or by
studies where a Polyakov-loop-extended Nambu Jona-Lasinio (PNJL) model
is directly matched to numerical results at imaginary $\mu$~\cite{takaha}.
Also various derivatives of the free energy with respect to the chemical potentials,
i.e. susceptibilities and generalized susceptibilities,
can be effectively computed by measuring the response of the thermal medium
to non-zero imaginary sources, then exploiting analytic continuation.
Systematic exploratory studies in this direction have been already 
performed~\cite{mdfs09}. In this context, one should also consider alternative 
ways for inserting the chemical potential on the lattice, see for instance
Ref.~\cite{gavai}, since that could change the effectiveness of analytic 
continuation.

\section{Electromagnetic background fields}

In the recent past
the possibility of performing standard Monte-Carlo simulations, together
with the interest related to various
theoretical and phenomenological issues, has fostered 
many lattice investigations of the non-perturbative properties of 
strong interactions in the presence of a magnetic background field 
$\mathbf B$.
While the initial primary interest, in connection with the phenomenology of 
non-central heavy ion collisions,
was in the so-called Chiral Magnetic Effect (CME)~\cite{cme}, i.e. the electric charge separation 
taking place along $\mathbf B$ in the presence of local CP-violating 
fluctuations, various other topics have been explored, going 
from magnetic catalysis, i.e. the expected enhancement 
of chiral symmetry breaking induced by $\mathbf B$~\cite{catalysis}, to the influence
of magnetic fields on the QCD phase diagram and the 
magnetic properties of strongly interacting matter. 
Some studies have provided unexpected results, which have 
fostered further investigations, both at the level 
of lattice QCD simulations and at that of theoretical models.

Magnetic catalysis of the QCD vacuum, i.e. the increase of the chiral
condensate as a function of $\mathbf B$, is a general phenomenon
which has been confirmed by many investigations, adopting different
discretizations and number of colors~\cite{buiv-catal,denecond,imps1,regencond,imps2}.
Instead, less expected developments have taken place at finite $T$.
Early determinations~\cite{dms10,imps1} of the dependence of the pseudo-critical 
temperature $T_c$ on $\mathbf B$, obtained by a standard
staggered discretization with unphysical quark masses, 
showed a weak increasing behavior of $T_c$
with $\mathbf B$, which was also in agreement with 
most model predictions. However, later simulations, obtained
with physical quark masses and closer to the continuum limit~\cite{rwdiagram},
have shown a decreasing behavior of $T_c$, which goes down by
 about 20\% for $eB \sim 1$ GeV$^2$; this is typically accompanied, in a region
around $T_c$, by a decrease
of the chiral condensate as a function of $B$, a phenomenon
known as inverse magnetic catalysis. Some evidence 
for inverse magnetic catalysis has been obtained also by simulations
of two-color QCD~\cite{imps2} and by
adopting dynamical overlap fermions~\cite{buivido}. 
A similar behavior ($T_c$ decreasing with $B$) was observed in the past 
in the case of chromo-magnetic backgrounds~\cite{bari}.

Regarding the chiral magnetic effect, lattice QCD studies
have focused on various different aspects, like: {\em i)} looking directly at electric charge separation
in a fixed topological and magnetic background~\cite{blum-cme};
{\em ii)} looking at the enhancement of longitudinal electric current fluctuations
in a magnetic background~\cite{buiv-cme}; {\em iii)} determining the longitudinal electric current
in the presence of a magnetic background and an axial chemical potential~\cite{yama-cme};
{\em iv)} measuring the 
correlation between electric polarization and topological charge 
in the presence of a 
magnetic background~\cite{rw-cme}. In general, results show a
suppression of the effect by about one order of magnitude
with respect to model predictions (a similar 
suppression is observed also for the so-called axial magnetic effect~\cite{braguta-cme}).
However, lattice systematics, in particular regarding a correct discretization
of chiral fermions, should be further checked in the future.

Many efforts have been dedicated to investigate the magnetic response
of strongly interacting matter, i.e. the dependence of its
free energy on the magnetic background, 
which reveals whether the material is paramagnetic or 
diamagnetic. 
This is, in principle, a clear-cut question for lattice QCD simulations,
since it regards the determination of purely 
equilibrium thermodynamical properties, contrary 
to other properties
like the electrical conductivity of strongly interacting matter~\cite{conducib}.
Nevertheless, obtaining a definite answer has been non-trivial
since in lattice simulations, which usually adopts
toroidal (hence compact) geometries in the spatial directions, 
the magnetic field is quantized, so that 
taking free energy derivatives with respect to $B$ 
is not a well defined operation.
After some first attempts, aimed at determining
only the spin component of the magnetization and suggesting
diamagnetism at all temperatures~\cite{regensusc0}, 
various improved methods have been devised. In particular one can: {\em i)} 
make use of 
thermodynamic integration in an extended parameter space
in order to obtain directly free energy differences, instead of
computing free energy derivatives~\cite{pisasusc0,pisasusc,regensusc3}; {\em ii)}
compute the magnetization in terms of pressure anisotropies in the 
direction parallel or orthogonal to $\mathbf B$~\cite{regensusc1,regensusc2};
{\em iii)} adopt an external field distribution which is not uniform 
and has zero net flux across the lattice torus, thus avoiding field quantization and allowing
for a direct computation of derivatives~\cite{levkova}.
Consistent results have been obtained by these methods and by different groups, showing that:
{\em i)} strongly interacting matter is strongly paramagnetic 
above deconfinement~\cite{pisasusc0,pisasusc,regensusc3,regensusc2,levkova}, with a response which is linear for magnetic
fields going up to $eB\sim 0.1-0.2$ GeV$^2$~\cite{pisasusc0,pisasusc} and a linear susceptibility
which grows logaritmically with $T$ in the high-$T$ regime~\cite{pisasusc0,pisasusc,regensusc3}; 
{\em ii)} the magnetic activity is instead very weak below $T_c$;
{\em iii)} the QCD vacuum itself has a non-linear paramagnetic 
behavior~\cite{regensusc1}. A strong paramagnetic behavior in the
deconfined phase is also the outcome of many model 
computations~\cite{suscmodels,endrodihrg}.

\begin{figure} 
\begin{center}
\includegraphics*[width=0.62\textwidth]{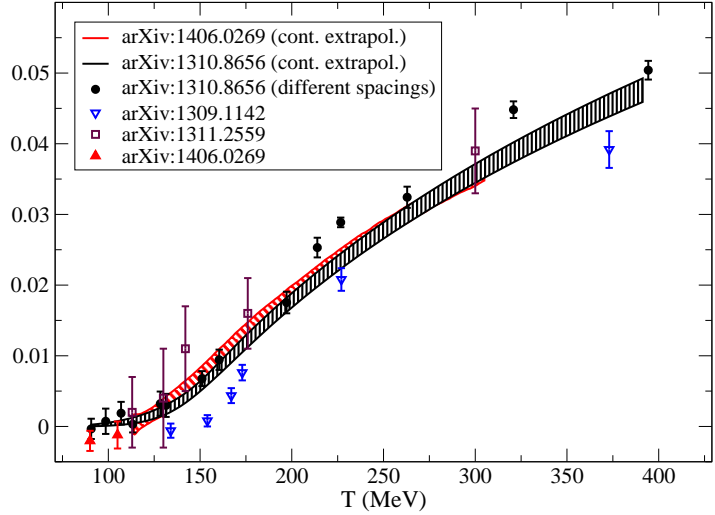}
\caption{Various determinations of the magnetic susceptibility of strongly 
interacting matter as a function of $T$ from studies at or close to
the physical point. Results from arXiv:1309.1142~\cite{levkova} (HISQ staggered 
quarks) adopt a direct determination of the free energy derivatives
in the presence of inhomogeneous magnetic fields; 
results from arXiv:1310.8656~\cite{pisasusc} (stout staggered fermions)
adopt thermodynamic integration, interpolating in $B$; 
results from arXiv:1311.2559~\cite{regensusc2} (stout staggered quarks) 
have been obtained 
by determining the magnetization via pressure anisotropies; 
finally, results from 
arXiv:1406.0269~\cite{regensusc3} (stout staggered quarks) adopt thermodynamic integration,
interpolating in the quark condensate.
}
\label{magnetisusc} 
\end{center}
\end{figure}

In Fig.~\ref{magnetisusc} we report a summary of the finite $T$ results,
which include all studies performed at or close to the physical
point (hence excluding early determinations with 
$N_f = 2$ unimproved staggered quarks~\cite{pisasusc0}, which however
provided qualitatively similar results), 
and are extrapolated to the continuum limit in two cases.
The agreement is especially good when results from 
Refs.~\cite{pisasusc,regensusc3,regensusc2} are compared: this 
is expected, since all three determinations adopt the same
discretization of the theory (stout staggered quarks, while 
Ref.~\cite{levkova} adopts HISQ staggered quarks), hence the comparison is a 
nice test that different methods to determine 
the magnetic susceptibility provide consistent results.
A few issues remain open, regarding in particular
the computation of 
non-linear corrections to the free-energy dependence of $B$, 
which could give important contributions to 
the equation of state in the regime of large fields, relevant to 
the early stages of the Universe, and the clarification
of the magnetic properties below $T_c$. Indeed, in a regime where
the thermal medium can be well approximated by a purely pion gas, computations based
on the Hadron Resonance Gas (HRG) model~\cite{endrodihrg} lead to the 
prediction of weak diamagnetic behavior~\cite{pisasusc}:
evidence for such a behavior at temperatures around 100 MeV is still
marginal~\cite{regensusc3}, while it clearly emerges when an isospin chemical potential is added
to the system, inducing pion condensation~\cite{endrodiso}.

Lattice simulations 
have also provided substantial evidence for  
a direct influence of magnetic fields on the gluon sector, mediated by
quark loop effects, which is expected due to the strongly
interacting nature of QCD in the low energy regime. 
Large part of magnetic catalysis
in the QCD vacuum is due to dynamical (sea) quark effects~\cite{denecond}, 
which are also thought to be at the origin of inverse magnetic catalysis~\cite{invcat,imps2}.
Magnetic field induced modifications and anisotropies are also visible in 
observables like the gluon action density~\cite{imps1,regensusc1} or the 
Polyakov loop~\cite{invcat}. 
Recently, it has been found that even the static quark-antiquark potential
undergoes significat modifications
in the presence of a magnetic background~\cite{anisotpot}: the string tension 
becomes weaker/stronger in the direction parallel/orthogonal to $\mathbf B$,
while the inverse occurs for the Coulomb coupling; such a behavior 
is consistent with some model computations~\cite{anisotmodel}.

Many studies have approached the problem of determining how the magnetic
field influences the hadron spectrum.
The interest has been driven mostly by the prediction for 
a superconductive phase of strong interactions above a critical
magnetic field, which would be signalled by $\rho$ meson condensation~\cite{chernosup}.
Early numerical evidence about this phenomenon~\cite{chernolat} has not
been confirmed by later studies~\cite{yamam,luschevskaya}, however the 
issue is still open.
An open question is also that regarding the possible effects of the observed anisotropy
of the static quark-antiquark potential on heavy meson states, and 
if these are eventually detectable in heavy ion experiments.

\section{QCD under combined external probes: topological effects
in external fields}

In many situations of theoretical or phenomenological interest,
one considers two different external 
sources at the same time, in order to probe new aspects of the theory.
An illuminating example is furnished by the CME
itself, where the combination of a magnetic background
and of an effective local fluctuation of the $\theta$ parameter
 leads to the phenomenon of electric charge separation, to 
be eventually observed in heavy ion experiments.

Such combinations have been explored by lattice simulations in many contexts.
Ref.~\cite{yama-cme} has explored the combined effect of an axial chemical potential
and a magnetic background field in order to obtain a numerical estimate
of the CME effect. Ref.~\cite{eb} has considered a CP-violating combination of
electric and magnetic backgrounds, 
i.e. such that $\mathbf E \cdot \mathbf B \neq 0$, in order to determine the induced
effective $\theta$ term in the pure gauge sector, which is linked to the effective
QED-QCD interactions in the pseudoscalar channel: in this case  
one relies on a double analytic continuation, since both $\mathbf E$ and the induced
effective $\theta$ are purely imaginary. In Ref.~\cite{endrodiso}, an isospin
chemical potential has been introduced, together with a 
non-zero magnetic background, in order to explore the magnetic
properties of isospin unbalanced matter. Various other possibilities
could be explored along the same line: for instance, were it not 
for the presence of the sign problem, numerical simulation of dense, cold
baryon matter in a magnetic background could open the exploration
of interesting phenomena, like the De Haas - van Alphen effect, which 
could be relevant to heavy astrophysical objects.

A particularly interesting issue is the study of the combined
application of a non-zero $\theta$-term and of an electric field.
Indeed, 
a numerical non-perturbative determination of the dependence of the electric dipole moment (EDM) of the
neutron on $\theta$, at least to the leading linear order,
 $d_N(\theta) = d_N^{(1)} \theta + O(\theta^3)$,
would help putting an upper bound on $\theta$ based 
on the experimental upper bound on the EDM of the neutron.
This is presently done by estimating $d_N^{(1)}$ by purely dimensional arguments.
A possible way to determine $d_N(\theta)$ is to measure
the three point function of an electromagnetic current between two neutron sources
at non-zero $\theta$.
$ \langle N(\mathbf p_1, s_1) | V_\mu^{EM} | N(\mathbf p_0, s_0) \rangle_\theta$.
The sign problem emerging from a non-zero $\theta$ can be approached 
by reweighting
configurations sampled at $\theta = 0$~\cite{shinta1}.
An interesting alternative is to introduce an imaginary 
$\theta$ and then perform analytic continuation~\cite{aoki_1},
the imaginary $\theta$ being introduced, after an anomalous axial transformation,
 as a $\gamma_5$ mass term in the fermionic action;
a recent study~\cite{guo} adopting this strategy has led to the upper
bound  $|\theta | \lesssim 7.6\, \times\, 10^{-11}$.
A different strategy, which has been proposed time ago~\cite{shinta2},
is to introduce an external electric 
background and measure the EDM from the energy difference between 
states with different spins in the presence of a $\theta$ term:
$${\cal E}^\theta_+(E_z) - {\cal E}^\theta_-(E_z) = d_N^{(1)} \theta E_z 
+ O(\theta^3) \, .$$
From a computational point of view, that could be very convenient, 
since it requires the determination of 
a two-point instead of a three-point function.
However, one has to face two different sign problems at the same time,
emerging from the non-zero electric field and from the non-zero $\theta$.
A possibility could be to combine
imaginary $\theta = i \theta_I$
and imaginary electric fields $\mathbf E = i \mathbf E_I$, perform direct 
determinations of the two point function
$${\cal E}^{\theta_I}_+(E_{Iz}) - {\cal E}^{\theta_I}_-(E_{Iz}) \simeq 
(i)^2 d_N^{(1)} \theta_I E_{Iz} =
- d_N^{(1)} \theta_I E_{Iz} 
$$ 
and then rely on a double analytic continuation in the two different imaginary sources:
preliminary results obtained by this approach have been reported in Ref.~\cite{takuedm}.

\section{Conclusions}

The insertion of external sources or background fields
is a common way to probe the properties of Quantum Field Theories.
In the context of lattice QCD simulations, one needs that the 
positivity of the path-integral measure be preserved by the 
insertion, otherwise standard Monte-Carlo methods cannot be used (sign problem).

In the case of magnetic background fields, no such problem appears,
so that lattice QCD simulations have been able to provide, in the last
few years, plenty of information about the QCD phase diagram in a
magnetic background and the magnetic properties of strongly interacting matter
at zero and finite $T$. Peculiar phenomena have also been found, like
inverse magnetic catalysis or magnetic field induced anisotropies
 in pure gauge quantities,
which claim for further investigations in the next few years.

In other cases, which include quark 
chemical potentials, electric background fields
and the $\theta$-term, 
the insertion makes the path-integral measure complex. 
A possible way out, in these cases, is to adopt 
purely imaginary sources, even if that does
not correspond to the physical interesting case.
One then typically relies on analytic continuation
in order to extract results for the real case,
however,  the information obtained
in the presence of imaginary sources can be interesting
by itself or for direct comparison with other models
and effective field theories.

One of the advantages of performing numerical simulations
at non-zero external sources, with respect to the standard
computation of expectation values at zero source in a Taylor
expansion approach, is that one typically needs correlations
functions and cumulants of lower order, with a great benefit
in terms of noise and of computational effort. 
The trade-off is given by the fact that one has to perform 
dedicated simulations at non-zero sources, however there is 
still room for various improvements that should be 
explored in the future.

In this review we have  discussed only a small part of the progress made in the
field in the recent past. In particular, we could not cover the
large efforts which are being made
in order to perform direct simulations in the presence of 
sources which make the path-integral measure complex,
i.e. in order to  provide real, instead of imaginary, solutions
to the sign problem.

\end{document}